\documentclass[prl,reprint,superscriptaddress,nofootinbib,twocolumn]{revtex4-2}

\usepackage[linktoc=page,colorlinks,urlcolor=blue,citecolor=blue,linkcolor=blue]{hyperref}
\usepackage{amssymb,amsmath}
\usepackage{graphicx}
\usepackage{natbib}
\usepackage{mathrsfs}
\usepackage{overpic}
\usepackage{gensymb}
\usepackage{color}
\usepackage{upgreek}
\usepackage[dvipsnames]{xcolor}
\usepackage[letterpaper,textwidth=7in,top=.75in,bottom=.75in]{geometry}
\usepackage{textcomp}

\usepackage{pdfpages}
\makeatletter
\AtBeginDocument{\let\LS@rot\@undefined}
\makeatother

\newcommand{\rload}{$\mathtt{R_{load}}$}

\newcommand{\snl}{Sandia National Laboratories, Albuquerque, New Mexico 87185, USA}
\newcommand{\cint}{Center for Integrated Nanotechnologies, Sandia National Laboratories, Albuquerque, New Mexico 87123, USA}

\begin{document}

\title{Measurement and Simulation  of the Magnetic Fields from a 555 Timer Integrated Circuit using a Quantum Diamond Microscope and Finite Element Analysis}
\date{\today}

\author{P. Kehayias}
\email{pmkehay@sandia.gov}
\affiliation{\snl}

\author{E. V. Levine}
\affiliation{The MITRE Corporation, Bedford, Massachusetts 01730, USA}
\affiliation{Quantum Technology Center, University of Maryland, College Park, MD 20742, USA}
\affiliation{Department of Physics, Harvard University, Cambridge, MA 02138, USA}

\author{L. Basso}
\affiliation{\cint}

\author{J. Henshaw}
\affiliation{\cint}

\author{M. Saleh Ziabari}
\affiliation{\cint}
\affiliation{University of New Mexico Department of Physics and Astronomy, Albuquerque, New Mexico 87131}

\author{M. Titze}
\affiliation{\snl}

\author{R. Haltli}
\affiliation{\snl}

\author{J. Okoro}
\affiliation{\snl}

\author{D. R. Tibbetts}
\affiliation{\snl}

\author{D. M. Udoni}
\affiliation{\snl}

\author{E. Bielejec}
\affiliation{\snl}

\author{M. P. Lilly}
\affiliation{\cint}

\author{T.-M. Lu}
\affiliation{\snl}
\affiliation{\cint}

\author{P. D. D. Schwindt}
\affiliation{\snl}

\author{A. M. Mounce}
\affiliation{\cint}

\begin{abstract}
Quantum Diamond Microscope (QDM) magnetic field imaging is an emerging  interrogation and diagnostic technique for integrated circuits (ICs). To date, the ICs measured with a QDM were either too complex for us to predict the expected magnetic fields and benchmark the QDM performance, or were too simple to be relevant to the IC community. In this paper, we establish a 555 timer IC as a ``model system" to optimize QDM measurement implementation, benchmark performance, and assess IC device functionality.  To validate the magnetic field images taken with a QDM, we used a SPICE electronic circuit simulator and Finite Element Analysis (FEA) to model the magnetic fields from the 555 die for two functional states. We compare the advantages and the results of three IC-diamond measurement methods, confirm that the measured and simulated magnetic images are consistent, identify the magnetic signatures of current paths within the device, and discuss using this model system to advance QDM magnetic imaging as an IC diagnostic tool.

\end{abstract}
\maketitle

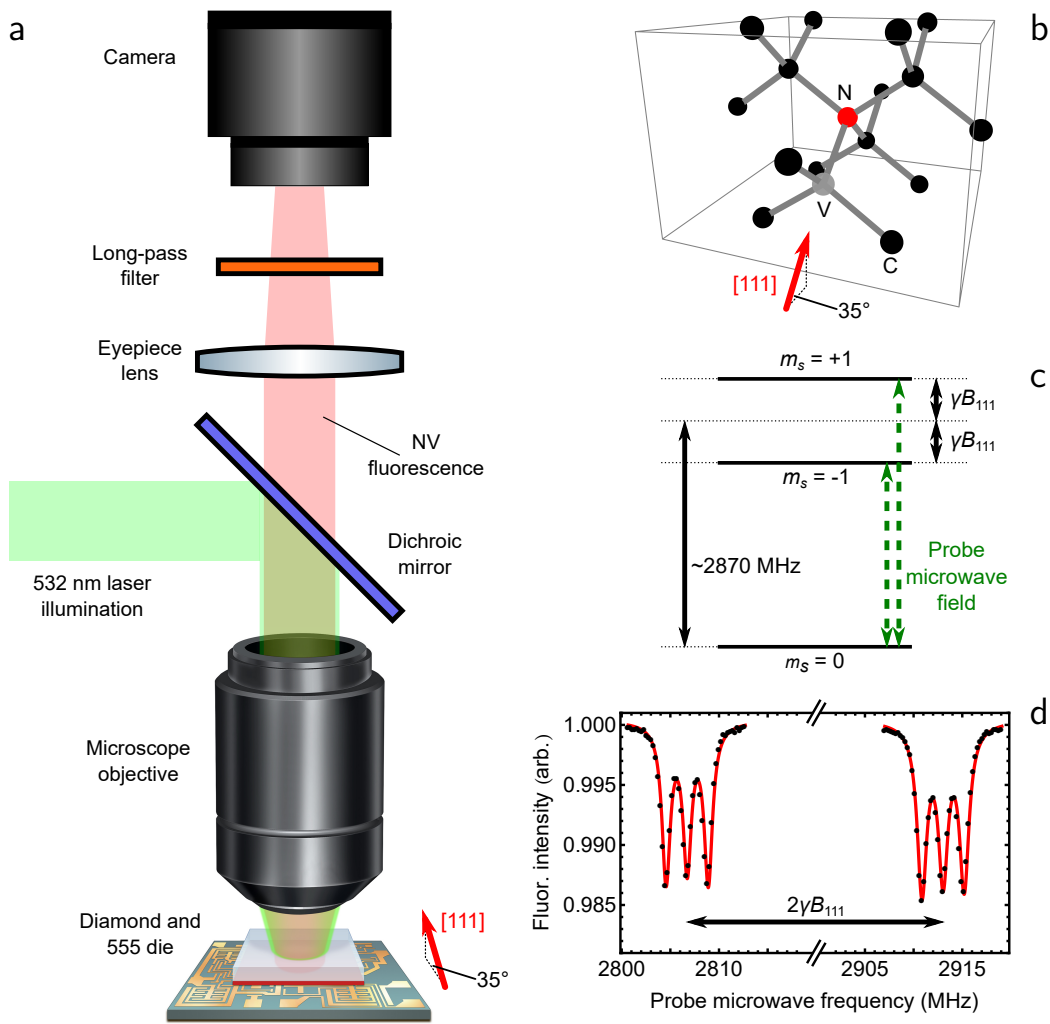
\begin{figure*}[ht]
\begin{center}
\begin{overpic}[width=0.75\textwidth]{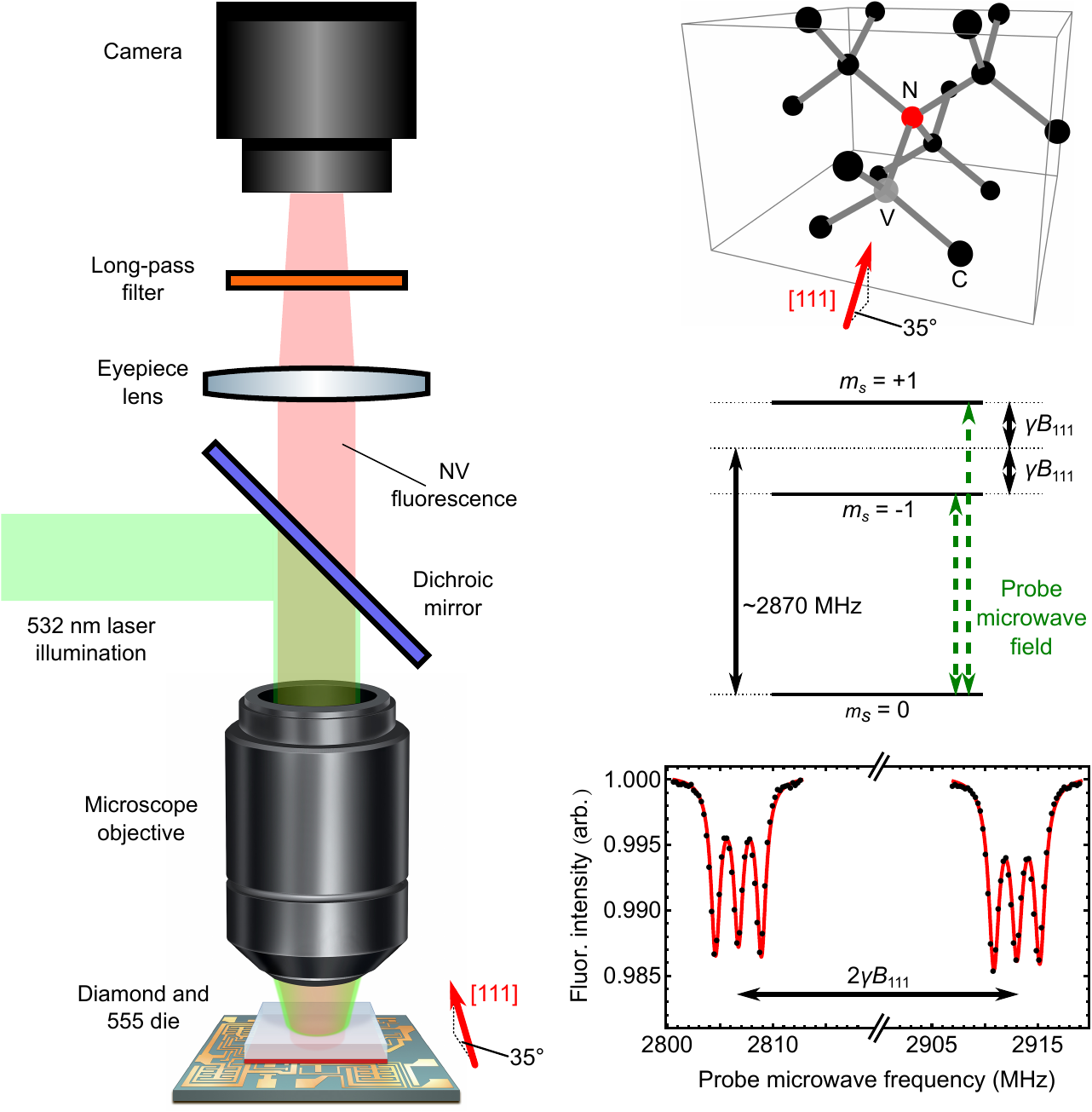}
\put(0,97){\textsf{\Large a}}
\put(100,97){\textsf{\Large b}}
\put(100,63){\textsf{\Large c}}
\put(100,30){\textsf{\Large d}}
\end{overpic}
\end{center}
\caption{\label{nvMagOverview}
(a) Schematic of an NV epifluorescence microscope setup measuring a 555 die (microwave wire and bias $B_{111}$ magnetic field not shown).
(b) NV center in the diamond lattice, with the arrow showing the diamond [111] direction.
(c) Energy level diagram of the NV ground-state magnetic sublevels, indicating the Zeeman effect and the zero-field splitting ($\sim$2870 MHz).
(d) Example ODMR spectrum for diamond Sample A. Each lineshape is split into three peaks due to the $^{14}$N hyperfine interaction.
}
\end{figure*}

\section{Introduction}

Mapping the magnetic fields from electric current distributions in an integrated circuit (IC) is a powerful noninvasive probing technique. Magnetic fields can provide information about the IC components, layout, and materials, as well as device function, fault locations, secure information leakage, and possible malicious hardware modifications (Trojans or counterfeits) \cite{orozcoBookCh, hwTrojanReview, counterfeitsReview}. Advances in device fabrication and packaging technologies have increased the IC complexity, requiring diagnostic techniques that can image devices with weaker electric currents and denser layouts, including devices with multiple conducting layers and 3D die packaging.

The Quantum Diamond Microscope (QDM) is emerging as a promising  IC diagnostic tool \cite{edlynQDMreview, QDM1ggg, degenReview}, allowing for non-destructive, high-resolution, wide-area magnetic field imaging of devices that is an alternative to scanning techniques such as superconducting quantum interference device (SQUID) microscopy, scanning giant magnetoresistance (GMR) microscopy, magnetic force microscopy (MFM),  or scanning magnetic tunnel junction (MTJ) microscopy \cite{felt_SQUID_MR, neoceraShorts, MFMshorts2, MTJshorts}. The QDM uses a layer of magnetically-sensitive nitrogen-vacancy (NV) color centers in diamond to image the magnetic fields from a nearby IC die \cite{edlynQDMreview}. QDM magnetic field imaging has been used to measure state-dependent magnetic fingerprints of a field-programmable gate array (FPGA), backside imaging of a flip-chip device, and hardware Trojan detection \cite{qdmFPGA, fpgaBackThin, qdmTrojan}. Advancement of the QDM technique as an IC diagnostic tool will benefit from a well-characterized system for benchmarking and optimizing sensor performance. To date, systems studied by the QDM have either been too simple indicate how the QDM will perform in operational setting with ICs \cite{NowodzinskiICmeas, OSAmicrowireQDM} or too complex to model the detected magnetic fields and the information they contain \cite{qdmFPGA, fpgaBackThin, qdmTrojan}. 

In this paper, we present experimental and computational results that map and simulate, respectively, the magnetic fields from a commercial bipolar junction transistor (BJT) 555 timer IC to benchmark and gauge QDM performance (such as magnetic sensitivity and spatial resolution). The 555 is an ideal ``model system" IC for QDM assessment, since it has $\sim$10-15 $\upmu$m features that are sufficiently coarse  to fully resolve, is simple enough to fully simulate, and is also among the most widely manufactured ICs \cite{designingAnalogChips}.  

We used a SPICE electronic circuit simulator (\texttt{PSPICE}) and  multiphysics Finite Element Analysis software (\texttt{COMSOL}) to simulate  the current densities and magnetic fields of the 555 die for two functional states. We measured the magnetic fields of these two states using a QDM, achieving micron-scale spatial resolution and few-$\upmu$T magnetic sensitivity in a 1$\times$1 $\upmu$m$^2$ pixel after 1 s.  Comparing the measured and the simulated magnetic maps, we confirmed that the QDM measured the expected magnetic field distributions, and assessed the QDM performance for three IC-diamond measurement configurations.  Analyzing the measured and simulated magnetic maps, we identified key features and current paths in the magnetic maps that correspond to the operational subsystems and internal states of the IC, showing how  QDM measurements can characterize the current densities and internal-state information of an IC. Finally, we discuss how our results apply to current mapping and failure analysis in other devices.

\begin{figure*}[ht]
\begin{center}
\begin{overpic}[width=0.95\textwidth]{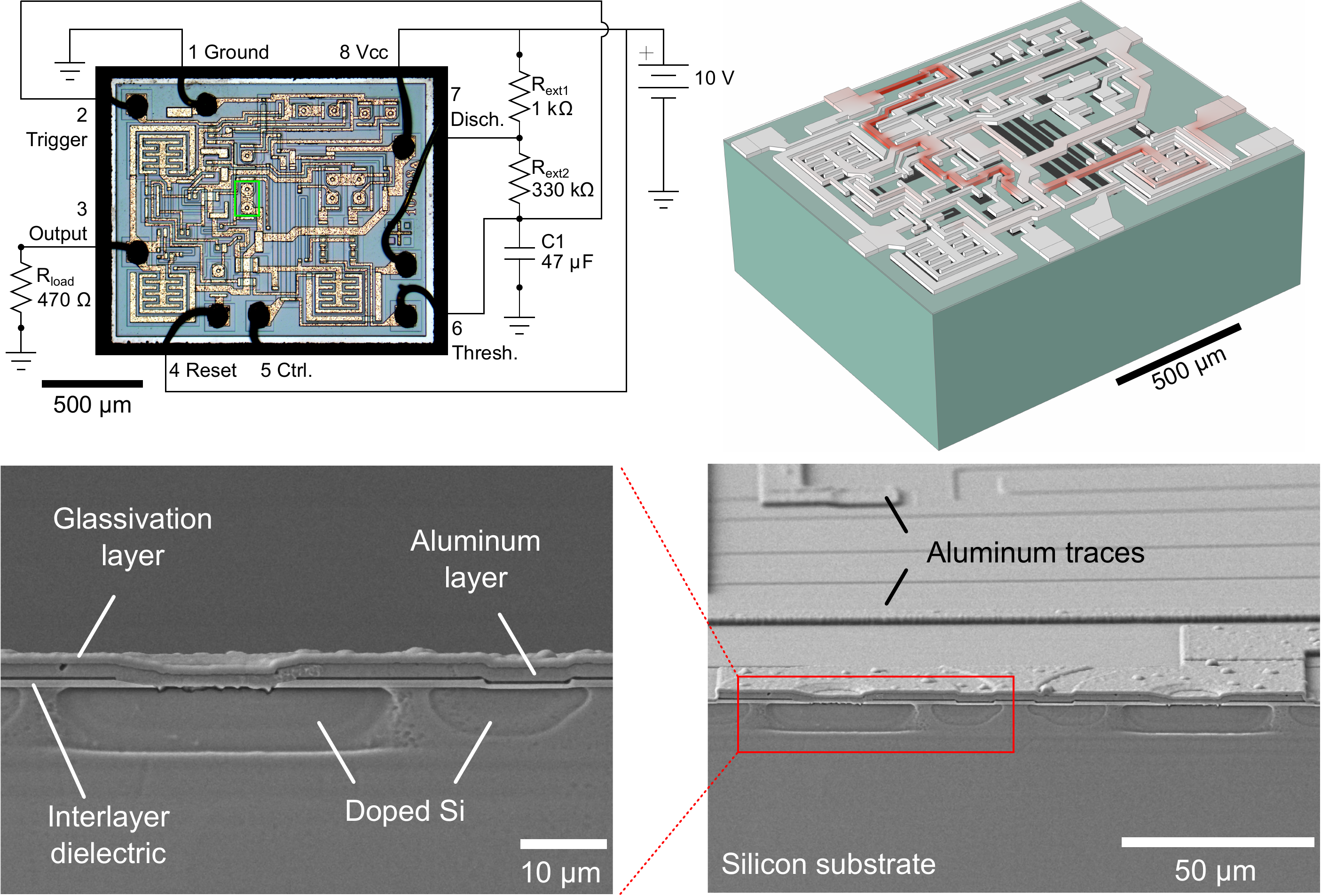}
\put(-1,68){\textsf{\Large a}}
\put(90,65){\textsf{\Large b}}
\put(-1,33){\textsf{\Large d}}
\put(53,33){\textsf{\Large c}}
\end{overpic}
\end{center}
\caption{\label{555dieImg}
(a) White-light optical microscope image of a 555 die, together with external components used to implement a two-state oscillator. The green box indicates area for cross-section SEM for subfigures (c)-(d).
(b) \texttt{COMSOL} model geometry for simulating a 555 die (silver, black, and green for aluminum, doped silicon, and substrate silicon, respectively), showing the simulated current density for the output-off state. 
(c) SEM image of a contact between the aluminum layer and the doped-silicon layers.
(d) Zoom of subfigure (c), which shows details of the metal and doped-silicon layers, which are insulated by a glassivation layer and an interlayer dielectric.
}
\end{figure*}

\begin{figure*}[htb]
\begin{center}
\begin{overpic}[width=0.90\textwidth]{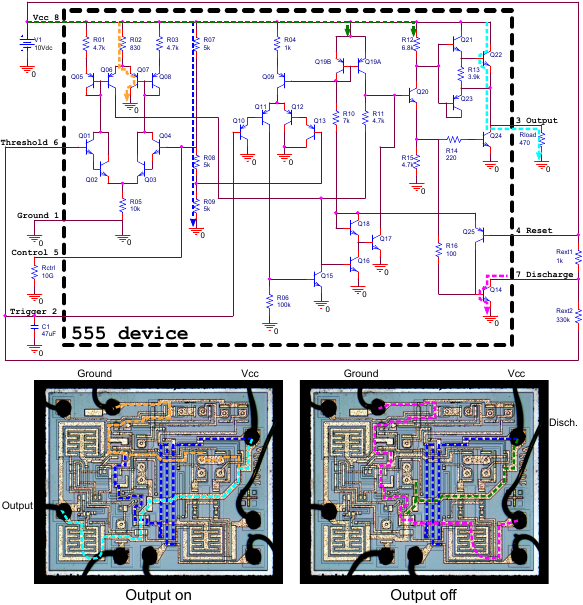}
\put(0,100){\textsf{\Large a}}
\put(2.5,36){\textsf{\Large b}}
\put(92,36){\textsf{\Large c}}
\end{overpic}
\end{center}
\caption{\label{currentPaths}
(a) Schematic for the 555 device and external components used to simulate currents in the two-state oscillator demo circuit. We implemented $\mathtt{Q19A}$ and $\mathtt{Q19B}$ as the two-collector BJT in the manufacturer's schematic, and used BJTs with two leads shorted as the diodes \cite{RCA555datasheet}. 
(b) For the output-on state, the cyan line indicates current to  \rload~and the orange line indicates current through $\mathtt{R02}$ and $\mathtt{Q07}$. 
(c) For the output-off state, the magenta line indicates the capacitor discharge current through $\mathtt{Q14}$ and the green line shows the current draw to the flip-flop and output networks. In both states, the dark blue line shows the current through three 5 k$\Omega$ resistors in series. All schematic components are labeled on the die photo in the supplemental material \cite{suppl}. 
}
\end{figure*}

\section{Devices and Materials}

\subsection{Experimental setup}

In each measurement, we placed a diamond sample on the 555 die NV-side down, then placed both in a fluorescence microscope (example apparatus  shown in Fig.~\ref{nvMagOverview}a). The diamond was illuminated with  532 nm laser light either from an angle (side illumination \cite{edlynQDMreview}) or through the microscope objective (epifluorescence illumination).  In the presence of a magnetic field $B_{111}$ along the N-V axis (the $[111]$ crystallographic direction, which is $\sim$35$\degree$ from the diamond surface), the resonance frequencies between the  $m_s = \pm 1$ ground-state magnetic  sublevels of NVs aligned along the $[111]$ direction are shifted by $\pm \gamma B_{111}$, where $\gamma \approx 28$ GHz/T is the NV gyromagnetic ratio (Fig.~\ref{nvMagOverview}b-c). We performed optically-detected magnetic resonance (ODMR) spectroscopy by driving microwave transitions between the  $m_s = 0$ and $m_s = \pm 1$ sublevels, which reduces the NV fluorescence intensity when the microwave frequencies are on resonance. Imaging the NV fluorescence intensity over a range of probe microwave frequencies, we obtain an ODMR spectrum for every pixel in camera's the field of view (Fig.~\ref{nvMagOverview}d).  We fit the ODMR spectrum in each pixel to extract the frequency splitting between the $m_s = 0$ to $\pm 1$ transitions, from which we generate a map of $B_{111}$.  We applied a static 1.5-2.5 mT bias magnetic field along the N-V axis, and we subtracted the magnetic field maps taken with and without the 555 energized to remove any contributions from sources other than currents in the 555 device.

\subsection{NV diamond samples}

Our magnetic imaging sensors are two single-crystal diamond samples with shallow NV surface layers. Both samples (Samples A and B) are 4$\times$4$\times$0.5 mm$^3$ electronic-grade diamond substrates with $<$5 ppb nitrogen density. One surface of Sample A has a 4 $\upmu$m $^{12}$C-enriched diamond layer with 20 ppm of $^{14}$N. NV centers were then formed using electron irradiation and vacuum annealing \cite{chu_annealing}.

The NV layer in Sample B was created by broad-beam $^{15}$N ion implantation  with 19 energies to form a uniform 1  $\upmu$m 50 ppm nitrogen layer \cite{srimFit, suppl}. After vacuum-annealing to activate NV formation, Sample B was laser-cut  into smaller pieces (1.14$\times$0.84$\times$0.5 mm$^3$) to match the 555 die dimensions for two of the IC-diamond integration methods, described below.  The surfaces of both diamonds were prepared by triacid cleaning (sulfuric, nitric, and perchloric), after which we coated the NV surfaces  with 5 nm of Ti to provide adhesion, 150 nm of Ag to prevent photoexcitation of electron-hole pairs in the device, and 150 nm of Al$_2$O$_3$ to prevent shorting between conductive elements.

\subsection{555 timer IC}

The 555 timer circuit was designed in 1971 using a BJT architecture, and it quickly became a best-selling IC used for a wide range of applications  \cite{designingAnalogChips}. The original design is largely unchanged except for an updated version that uses complementary metal–oxide–semiconductor (CMOS) technology, which requires less current for operation, while the BJT version allows for larger current throughput. Here, we studied the RCA CA555CE BJT 555 timer \cite{RCA555datasheet}, a BJT version of the IC with $\sim$10-15 $\upmu$m features that, when carrying current, are sufficiently large to magnetically detect and resolve with our QDM.

The 555 timer die (Fig.~\ref{555dieImg}) has two conducting layers: a $\sim$1.6 $\upmu$m top aluminum layer and a $\sim$6.4 $\upmu$m doped-silicon layer, separated by an interlayer dielectric with contacts between layers. We determined the layer thicknesses by cross-sectioning a die that was removed from its eight-pin dual in-line package (DIP). We chemically stained the cross section to reveal the doped silicon layer, and imaged using scanning electron microscopy \cite{suppl}. The IC consists of NPN junctions,  PNP junctions, and doped-silicon resistors arranged into Darlington pairs, current mirrors, voltage comparators, a voltage divider,  and a flip-flop (Fig.~\ref{currentPaths}). 

We configured the IC as a two-state oscillator by adding three external resistors and a capacitor ($\mathtt{R_{ext1}}$, $\mathtt{R_{ext2}}$, \rload, and $\mathtt{C1}$, shown in Fig.~\ref{555dieImg}a) \cite{RCA555datasheet, designingAnalogChips, mims555, digitalDesign, suppl}. In this circuit, the voltage across $\mathtt{C1}$ oscillates between $\mathtt{Vcc}$/3 and 2$\mathtt{Vcc}$/3. Depending on the $\mathtt{C1}$ voltage and the flip-flop state, the 555 device will either be in an output-on state (sourcing current to \rload) or an output-off state (discharging $\mathtt{C1}$ through the IC to ground). In Fig.~\ref{currentPaths} we indicate some of the primary current paths through the device in each internal state.  These current paths create visible features highlighted in the QDM magnetic images below, which shows how QDM measurements can collect the internal-state information of an IC.

The oscillation frequency $f$ and duty cycle $D$ are determined by  $\mathtt{R_{ext1}}$, $\mathtt{R_{ext2}}$, and $\mathtt{C1}$:
\begin{gather*}
  f = \frac{1}{ \ln{2} ~\left( \mathtt{R_{ext1}} + 2 \mathtt{R_{ext2}} \right) \mathtt{C1}} , \\
  D = \frac{ \mathtt{R_{ext2}} }{ \mathtt{R_{ext1}} + 2 \mathtt{R_{ext2}} } .
\end{gather*}
We chose  $\mathtt{R_{ext1}}$ = 1 k$\Omega$, $\mathtt{R_{ext2}}$ = 330 k$\Omega$, and $\mathtt{C1} = 47~\upmu$F to get a $\sim$20 s period and a 50\% duty cycle, confirmed with oscilloscope measurements to demonstrate that the device was in working condition  \cite{suppl}. During the magnetic imaging measurements, we kept the 555 in the output-on state by connecting  $\mathtt{C1}$ to ground, or in the output-off state by connecting $\mathtt{C1}$ to $\mathtt{Vcc}$.

\begin{figure*}[htb]
\begin{center}
\begin{overpic}[width=0.95\textwidth]{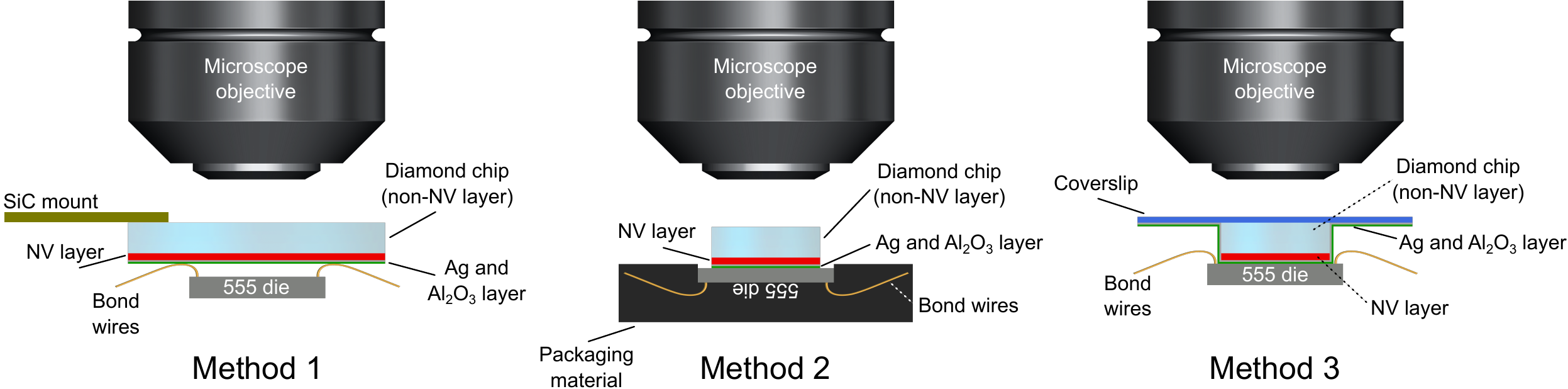}
\put(5,23){\textsf{\Large a}}
\put(37.5,23){\textsf{\Large b}}
\put(70,23){\textsf{\Large c}}
\end{overpic}
\end{center}
\caption{\label{microscopeDrawings}
Schematic drawings for the three IC-diamond integration methods we used: (a) diamond over wire bonds, (b) backside thinning, (c) and diamond between the wire bonds.
}
\end{figure*}

\begin{figure*}[htb]
\begin{center}
\begin{overpic}[width=0.85\textwidth]{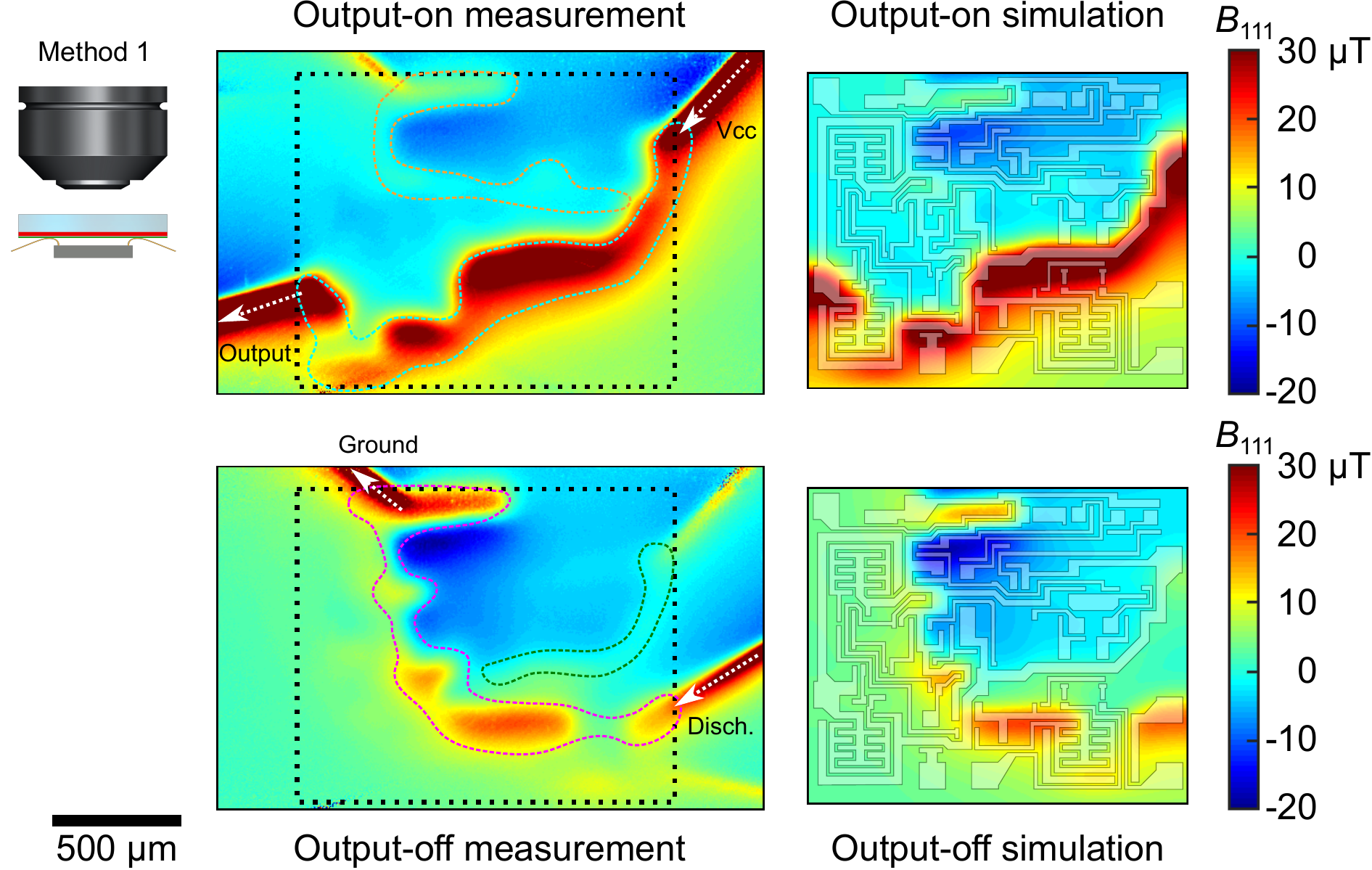}
\put(0,8){\includegraphics[width=0.12\textwidth]{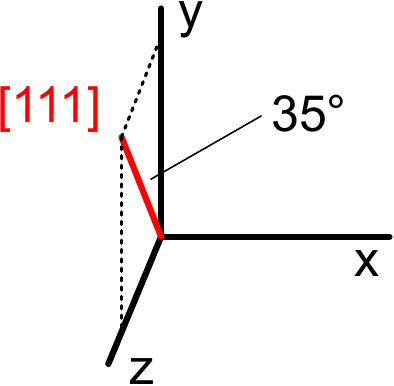}}
\end{overpic}
\end{center}
\caption{\label{bigDiamondTop}
\textit{Method 1: Diamond over wire bonds.}
Measurements (left) and simulations (right) of a 555 die in the output-on (top) and output-off (bottom) state. The simulation (and black dotted box) dimensions are 1216$\times$1464 $\upmu$m$^2$. This measurement has a $z_\textrm{fit} = 71~\upmu$m standoff distance from the die surface, and the simulated magnetic field maps displayed are calculated at $z = 71~\upmu$m for comparison. These measurements have a 0.3 $\upmu$T noise floor in a 1$\times$1 $\upmu$m$^2$ pixel. The white arrows  show the primary current input/output points for the device, and the circled regions show magnetic features from currents in Fig.~\ref{currentPaths} (with the same colors).
}
\end{figure*}

\begin{figure*}[htb]
\begin{center}
\begin{overpic}[width=0.85\textwidth]{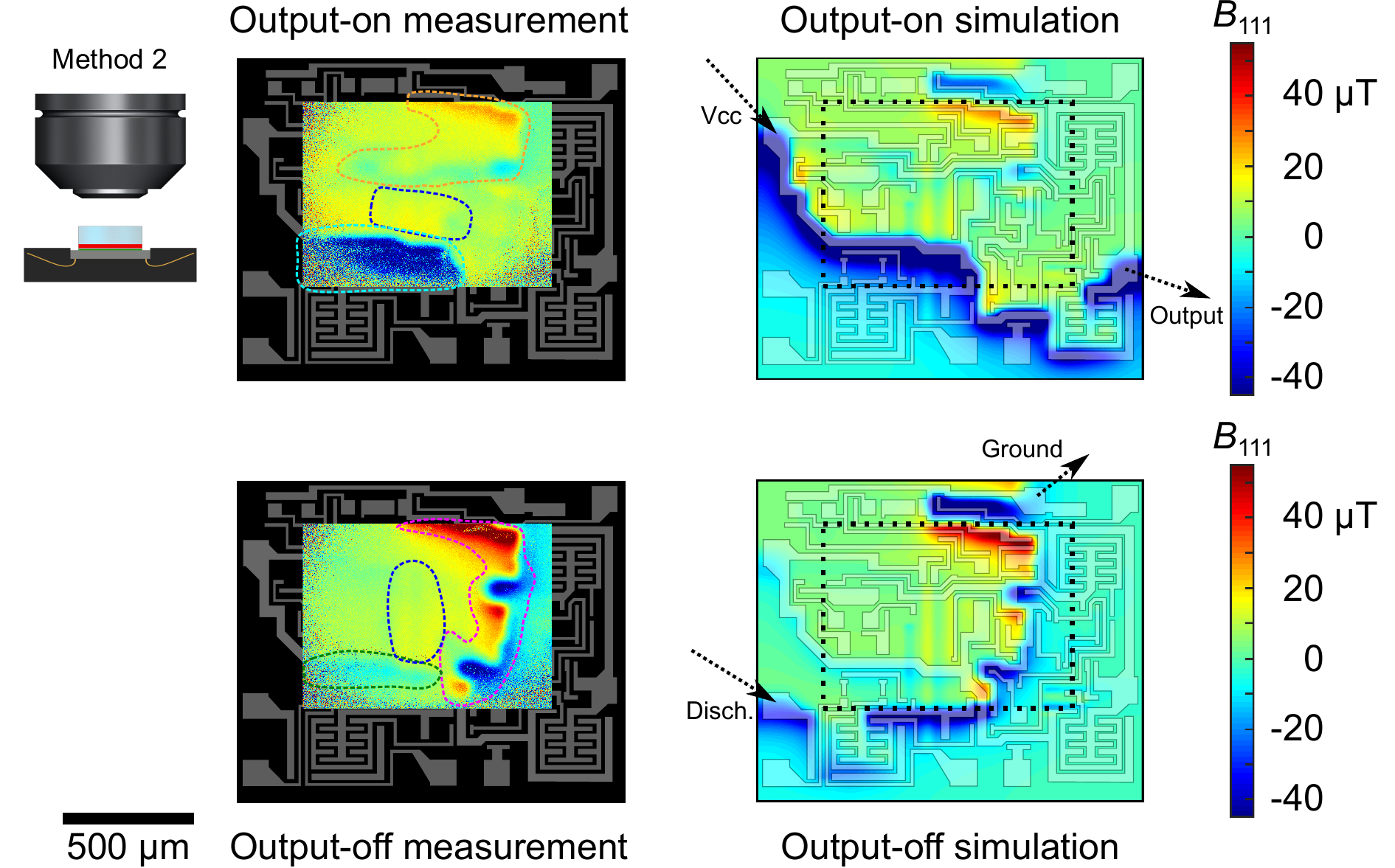}
\put(0,8){\includegraphics[width=0.12\textwidth]{b111Coords.pdf}}
\end{overpic}
\end{center}
\caption{\label{backThinned}
\textit{Method 2: Backside thinning.} 
Measurements (left) and simulations (right) of a 555 die in the output-on (top) and output-off (bottom) state. Note that the die is left-right flipped compared to the other images. This measurement has a $z_\textrm{fit} = 26~\upmu$m standoff distance (measured from the surface of the doped-silicon layer), and the simulated magnetic field maps displayed are calculated at $z = 26~\upmu$m for comparison. These measurements have a 3 $\upmu$T noise floor in a 1$\times$1 $\upmu$m$^2$ pixel. The black arrows  show the primary current input/output points for the device, and the circled regions show magnetic features from currents in Fig.~\ref{currentPaths} (with the same colors).
}
\end{figure*}

\begin{figure*}[htb]
\begin{center}
\begin{overpic}[width=0.85\textwidth]{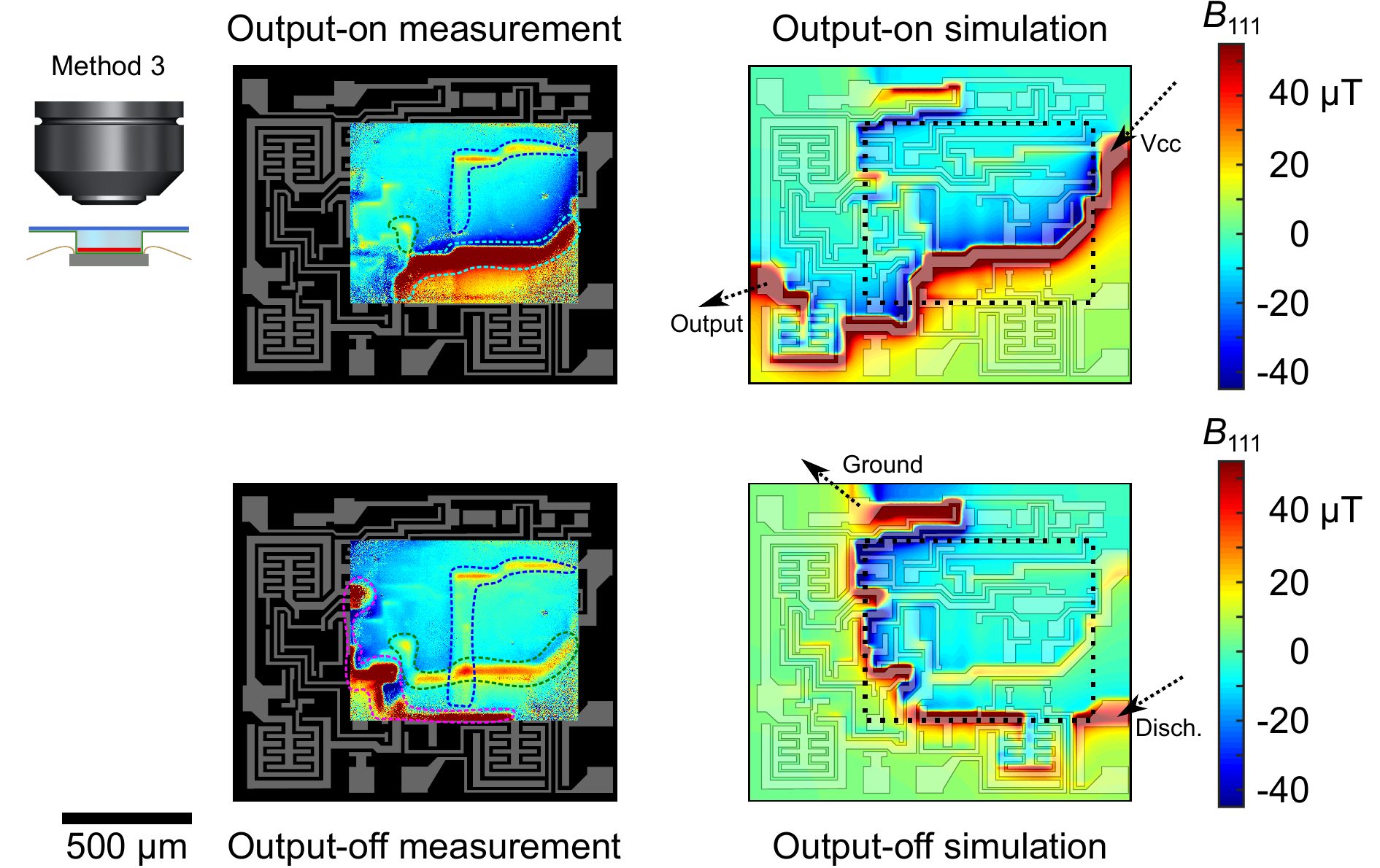}
\put(0,8){\includegraphics[width=0.12\textwidth]{b111Coords.pdf}}
\end{overpic}
\end{center}
\caption{\label{smallDiamondTop}
\textit{Method 3: Diamond between the wire bonds.} 
Measurements (left) and simulations (right) of a 555 die in the output-on (top) and output-off (bottom) state. This measurement has a $z_\textrm{fit} = 4~\upmu$m standoff distance from the die surface, and the simulated magnetic field maps displayed are calculated at $z = 4~\upmu$m for comparison. These measurements have a 3 $\upmu$T noise floor in a 1$\times$1 $\upmu$m$^2$ pixel. The black arrows  show the primary current input/output points for the device, and the circled regions show magnetic features from currents in Fig.~\ref{currentPaths} (with the same colors).
}
\end{figure*}

\section{IC-diamond integration methods}
We imaged magnetic fields from the 555 die using three IC-diamond integration methods: a top approach with a large diamond over the bond wires (Fig.~\ref{microscopeDrawings}a), a back approach with a small diamond in the slot of a back-thinned die (Fig.~\ref{microscopeDrawings}b), and a top approach with a small diamond between the  wire bonds (Fig.~\ref{microscopeDrawings}c), ordered from largest to smallest IC-diamond standoff distance. Each  approach uses a similar NV imaging apparatus, but they differ in the IC and diamond preparation steps. During each measurement, we monitored the voltage across \rload ~to ensure that there were no bent bond wires causing a short, light leakage causing photocurrent in the die, or other failure modes.

\subsection{Method 1: Diamond over wire bonds}
Here we used a diamond (Sample A) larger than the die to image the magnetic fields from the entire 555 die and the bond wires (Fig.~\ref{microscopeDrawings}a). To minimize the standoff distance between the NV layer and the die surface, we removed the die from its packaging by placing the IC in 90\% fuming nitric acid at 90 \degree C for 15 minutes. We then rinsed the fully-exposed die with acetone, isopropyl alcohol, and deionized water and removed the bond wires using tweezers. We then glued the die to an 8-pin DIP breakout board, electrically connected it with 0.001" diameter gold wedge wire bonds, and installed it into a perfboard with the external components.

We affixed Sample A in the QDM setup (glued to a protruding piece of silicon carbide) with the NV layer at the microscope focal plane, and illuminated the NV layer using side illumination.  We positioned the 555 die under the NV layer using a stepper-motor translation stage, measuring with decreasing standoff distances until the device stopped working properly. Using this large 4$\times$4$\times$0.5 mm$^3$ diamond sample facilitates mounting it in the microscope setup, but since it is larger than the 555 die, the standoff distance was limited by the bond wires touching the diamond surface.

\subsection{Method 2: Backside thinning} 
To prepare the 555 die for this approach, we first cut through the  back of the
packaging and the copper ground plane using an Allied X-Prep mill to gain access to the backside of the  die. We then thinned the exposed silicon to 20-30 $\upmu$m, which was the minimum thickness for which the die was still functional. The DIP pins were bent 180\degree, then connected to a perfboard with a mirror-flipped layout compared to the layout for Method 1.

We measured the magnetic map using a 1.14$\times$0.84$\times$0.5 mm$^3$ piece of Sample B placed on the bottom side of the die (Fig.~\ref{microscopeDrawings}b). To avoid shadows from the packaging and from the sides of diamond, we illuminated the NV layer  using epifluorescence illumination.

\subsection{Method 3: Diamond between the wire bonds}
For this method, we used the wedge-bonded die from Method 1, as well as die decapsulation. To expose the die while keeping it in the original packaging, we decapsulated a 555 DIP IC using an etch tool (RKD Mega Etch) with a custom gasket and fuming nitric acid (95 \degree C for 10 seconds to soften the polyimide, followed by 85 \degree C for 25 seconds).  This removed the packaging material to expose the die while maintaining functionality.

For the measurement, we used another laser-cut  piece of Sample B, and again used epifluorescence illumination. The main difference in this method is that we glued the diamond to a cover slip with UV-curing transparent glue (Fig.~\ref{microscopeDrawings}c).  Both the diamond and the cover-slip were coated with the Ti-Ag-Al$_2$O$_3$ adhesion-reflection-insulation layers to prevent light from leaking around the sides of the diamond.

\section{Computational Methods}

We developed a finite element simulation of the magnetic fields from the 555 die for the output-on and output-off states to evaluate the standoff distance and spatial resolution of each measurement, and to check that the measured magnetic field maps were consistent with what is expected from simulation.

\subsection{Governing equations}

For the output-on and output-off states, the magnetic fields generated by the 555 die depend on the current density distributions and the material characteristics of the device. The dynamics of the $\sim$20 s period two-state oscillator are slow enough for us to treat the output-on and output-off states independently in the static limit. The current density in the device is given by the continuity equation and Ohm's law:
\begin{equation}
\pmb{\nabla}\cdot\mathbf{J} = -\pmb{\nabla}\cdot \left(\sigma \pmb{\nabla} V\right)=0,
    \label{Continuity}
\end{equation}
where $\mathbf{J}$ is the current density in the 555, $\sigma$ is the electrical conductivity of the 555 materials, and $V$ is the electrical potential field resulting from the external circuit shown in Fig.~\ref{555dieImg}(a). The magnetic field is determined by the static Amp\`ere's law:
\begin{equation}
\pmb{\nabla}\times\frac{\pmb{\nabla}\times\mathbf{A}}{\mu} = \mathbf{J},
    \label{StaticAmpere}
\end{equation}
where $\mathbf{A}$ is the vector magnetic potential field and $\mu$ is the magnetic permeability of the 555 materials.

We sequentially solve the system of governing equations (Eqns.~\ref{Continuity} and \ref{StaticAmpere}) to determine the magnetic field maps at different standoff distances. We first solve Eqn.~\ref{Continuity} to determine the scalar electrical potential field, $V(\mathbf{r})$, with the appropriate boundary conditions for $V$. We use Ohm's law to determine $\mathbf{J}$ in the 555, which we then use to solve Eqn.~\ref{StaticAmpere} for $\mathbf{A}$ with appropriate boundary conditions. We then calculate the magnetic field, $\mathbf{B}(\mathbf{r})$, using $\mathbf{B}=\pmb{\nabla}\times\mathbf{A}$.

\subsection{555 geometry and boundary conditions}
The 555 die model consists of a conducting metal layer, a conducting doped silicon layer, and an interlayer dielectric to electrically insulate the aluminum from the doped silicon as shown in Fig.~\ref{555dieImg}d. Electrical contact between the two conducting layers occurs only at specific locations on the die to connect the internal device components of the 555.

We solved the system of governing equations  using the finite element software \texttt{COMSOL Multiphysics}$^\text{\textregistered}$ version 5.5, implementing the geometry and material properties of the 555. Finite element modeling allows us to include the detailed multi-layer geometric features of the 555 die with dimensions extracted  from the optical microscope image in Fig.~\ref{555dieImg}a. Fig.~\ref{555dieImg}b depicts the the full 3D model geometry built using \texttt{COMSOL}, with the top metal layer in silver, and the doped silicon layer in black. The un-doped silicon die is shown in green. The air domain above the die that was also included to complete the computational domain of the model is not shown. 

The boundary conditions for Eqn.~\ref{Continuity} are electrically insulating with zero normal current density everywhere except for the external boundaries corresponding to the eight pins of the device. These pins are connected to an external circuit and have voltage determined by the parameters of the externally connected circuit components in Fig.~\ref{555dieImg}a. The tangential components of the magnetic vector potential field $\mathbf{A}$ are set to zero for the boundary condition of Eqn.~\ref{StaticAmpere}.

\subsection{Material properties and SPICE circuit simulation}

We estimated the electrical conductivities ($\sigma$) and the magnetic permeabilities ($\mu$) of the 555 internal components from available information about the device. The top metal layer was assumed to be aluminum, based on the SEM analysis in Fig.~\ref{555dieImg}c-d. We modeled the interlayer dielectric using insulator properties of SiO$_2$. We used the magnetic permeability of silicon  for all domains in the silicon layer and estimated the electrical conductivity of the resistor and transistor elements. We computed resistor conductivities using the resistances reported in the 555 circuit schematic in the datasheet (Fig.~\ref{currentPaths}a) and the geometrical dimensions of the resistor region approximated from the optical images of the die.

To determine the current in each transistor, we simulated the 555 device in the two-state oscillator circuit in a SPICE electronic circuit simulator (\texttt{PSPICE}) by combining the manufacturer's schematic with the external circuit components (Fig.~\ref{currentPaths}a) \cite{RCA555datasheet}. We measured the 555 external voltages, currents, and dynamics (frequency and duty cycle) to confirm that the model was performing as expected \cite{suppl}, giving us confidence that this simulation also predicted the internal behavior correctly. The SPICE model provided information about the current in each transistor element, from which we estimated the electrical conductivities for the relevant parts of the die model.

\subsection{Numerical implementation and output}

We solved the governing equations using a steady-state model in \texttt{COMSOL}, adjusting the mesh resolution such that the solution remained constant when changing the spatial discretization parameters. A triangular mesh with a minimum element size approximately 6\% of the minimum geometrical feature size was required in the neighborhood of the metal, insulator, and semiconducting layers. We coupled the Electric Currents (\texttt{ec}) module with the Electrical Circuit (\texttt{cir}) module to solve the continuity equation (Eqn.~\ref{Continuity}) for $V(\mathbf{r})$. The external circuit elements in Fig.~\ref{555dieImg}(a), constructed using the \texttt{cir} module, set the boundary conditions of the finite element computational domain on the pins. The output, $V(\mathbf{r})$, from the coupled \texttt{ec} and \texttt{cir} modules, determines the current density $\mathbf{J}(\mathbf{r})$. We use this computed current density as the input to the Magnetic Fields (\texttt{mf}) module to solve Amp\`ere's law (Eqn.~\ref{StaticAmpere}) to determine $\mathbf{B}(\mathbf{r})$. The full magnetic field solution allowed us to calculate the $B_{111}(\mathbf{r})$ field component in planes of different standoff distance above and below the die, which we used in analysis and comparison with measurements.

\section{Results and Discussion}

Figures \ref{bigDiamondTop}-\ref{smallDiamondTop}  show the magnetic maps measured with each method, together with a simulated magnetic map for comparison.  We used the outputs from the \texttt{COMSOL} simulations to determine the standoff distance $z_\textrm{fit}$ for each measurement. To do this, we created an interpolating function $B_{111}(x, y, z)$ that is continuous in the spatial coordinate variables $\{x, y, z\}$, using a set of the simulated magnetic field maps generated by the \texttt{COMSOL} simulation over a range of altitudes $z$ (up to 100 $\upmu$m from the die surface). We then performed a least-squares fit between the measured $B_{111}$ and the simulated $B_{111}(x, y, z)$ fit function. The optimal $z_\textrm{fit}$ gives our NV-die standoff distance and the optimal $x_\textrm{fit}$ and $y_\textrm{fit}$ are spatial offsets between the measurement and the simulation. This fit was performed for both the output-on and output-off device states, and we found close agreement between the measured and the simulated magnetic maps. This analysis returned $z_\textrm{fit} = \{71,26,4\}$ $\upmu$m standoff distances for the three methods, with a $\{2,3,0.3\}$ $\upmu$m uncertainty for each.

Figure \ref{bigDiamondTop} shows the resulting magnetic field maps measured with Method 1 (diamond over wire bonds) at the closest possible standoff distance of 71 $\upmu$m. Since the diamond sample is fixed in the QDM apparatus and is large compared to the die, this approach allows for good heat sinking from pump laser heating, a large image field of view (including the entire die and the bond wires), good stray light protection, and variable-altitude measurements. However, the standoff distance is limited by the bond wires, which reduces the field strength and the spatial resolution. Furthermore, since the bond wires are touching the diamond, the fields measured from the bond wires are the stronger than those of the 555 die.

Figure \ref{backThinned} shows the magnetic field maps measured with Method 2 (backside thinning) at a 26 $\upmu$m standoff distance from the doped-silicon layer surface. These magnetic images look different compared to those of the other methods since we measure from the back of the die (flipped left-right compared to the other die images) but keep the same $B_{111}$ projection direction.  Method 2  benefits from being able to preferentially detect current in the doped-silicon layer with greater signal-to-noise ratio and spatial resolution.  However, the necessary die-thinning step may limit the  utility of this magnetic imaging implementation. Furthermore, due to the reduced die thickness and poor thermal conductivity of the packaging, the die has poor heat sinking from laser heating. This limits the maximum-allowable laser power and the magnetic sensitivity for this method.

Figure \ref{smallDiamondTop} shows the magnetic field maps measured with Method 3 (diamond between the wire bonds). Comparing to the simulated magnetic maps, we determined a 4 $\upmu$m standoff distance for these measurements. With this standoff distance we can image the weaker currents in the die, which are consistent with the fields predicted by \texttt{COMSOL}. This approach had the best standoff distance, spatial resolution, and field strength, and is also ideal for imaging dies where the wire bond spacing is not a limitation to the standoff distance \cite{qdmFPGA}. However, avoiding laser heating is more challenging than with Method 1, since heat from the diamond can flow to the environment through the silicon carbide mount in Method 1.

Each measurement method achieved a different standoff distance to the die, listed in Table \ref{schemesTable}. The spatial resolution of an NV magnetic imaging measurement is influenced by the optical diffraction limit ($\sim$1 $\upmu$m), the standoff distance, and the NV layer thickness \cite{edlynQDMreview}. In this work, the standoff distance (set by the bond wires, silicon thickness, or dust on the die surface) was the main limitation to the spatial resolution, and is also an estimate for the minimum spatial resolution for each measurement. As illustrated in Figs.~\ref{bigDiamondTop}-\ref{smallDiamondTop}, decreasing the standoff distance enhances the magnetic feature resolution in the 555 magnetic maps. Note that the feature sizes in the 555 die are coarse enough that measuring with a standoff distance smaller than 4 $\upmu$m would not reveal more detail.

Table \ref{schemesTable} also lists each magnetic noise floor $\delta B_{111}$, which is the standard deviation of the measured magnetic fields in the field of view when measuring with the 555 disconnected, after subtracting a background measurement. These $\delta B_{111}$ values are normalized to a 1$\times$1 $\upmu$m$^2$ pixel size and a 1 s experiment duration. Sample A has a better magnetic sensitivity than Sample B, due to its NV fluorescence contrast, resonance linewidth, and fluorescence intensity being better \cite{suppl}. To quote the projected best-case $\delta B_{111}$ if using Sample A, we also evaluated $\delta B_{111}$ for Method 2 and Method 3 when using Sample A in the same conditions. For comparison, Table \ref{schemesTable} also lists the maximum $B_{111}$ at each standoff distance, from which we can calculate a maximum signal-to-noise ratio (SNR) by dividing by $\delta B_{111}$.  Note that some experiments had an SNR $< 1$ while the images in Figs.~\ref{bigDiamondTop}-\ref{smallDiamondTop} had an SNR $> 1$ because of longer averaging times (typically 10 minutes to 1 hour).   The SNR improves with closer standoff distance, though the maximum $B_{111}$ improves slower than $1/z$ (as with current in an infinite straight wire) for $z \lesssim 25~\upmu$m because of the finite size of the conductors. We also convert the $\delta B_{111}$  and $z_\textrm{fit}$ to a current sensitivity $\delta I$ and a surface current density $\delta K$ in Table \ref{schemesTable} assuming an infinite wire or an infinite sheet of current along the $+x$ direction, respectively \cite{suppl}.  When measuring static currents, the reported $\delta I$ and  $\delta K$ sensitivities can be enhanced with coarser pixel size and longer experiment duration (compared to 1$\times$1 $\upmu$m$^2$ pixel size and 1 s experiment duration).

\begin{table*}[]
\begin{tabular}{c|c|c|c|c|c|c}
& \begin{tabular}[c]{@{}c@{}}\textbf{Standoff}\\ \textbf{distance $z_\textrm{fit}$}\end{tabular} & \begin{tabular}[c]{@{}c@{}}\textbf{$\delta B_{111}$}\\ \textbf{(1 s, 1$\times$1 $\upmu$m$^2$)}\end{tabular} & \begin{tabular}[c]{@{}c@{}}\textbf{Max  $B_{111}$}\\ \textbf{from device}\end{tabular} & \begin{tabular}[c]{@{}c@{}}\textbf{Max $B_{111}$ SNR}\\ \textbf{(1 s, 1$\times$1 $\upmu$m$^2$)} \end{tabular} & \begin{tabular}[c]{@{}c@{}}\textbf{$\delta I$}\\ \textbf{(1 s, 1$\times$1 $\upmu$m$^2$)}\end{tabular} & \begin{tabular}[c]{@{}c@{}}\textbf{$\delta K$}\\ \textbf{(1 s, 1$\times$1 $\upmu$m$^2$)}\end{tabular}  \\
\hline
Method 1 & 71 $\upmu$m    & 8.0 $\upmu$T & 42 uT & 5 & 3.1 mA & 16 A/m \\
\hline
Method 2 & 26 $\upmu$m    & \begin{tabular}[c]{@{}c@{}}160 $\upmu$T\\ (4.0 $\upmu$T)\end{tabular} & 98 $\upmu$T    & \begin{tabular}[c]{@{}c@{}}0.6\\ (25)\end{tabular} & \begin{tabular}[c]{@{}c@{}}23 mA\\ (560 $\upmu$A)\end{tabular}   & \begin{tabular}[c]{@{}c@{}}310 A/m\\ (7.7 A/m)\end{tabular} \\
\hline
Method 3 & 4 $\upmu$m     & \begin{tabular}[c]{@{}c@{}}100 $\upmu$T\\ (3.6 $\upmu$T)\end{tabular} & 220 $\upmu$T   & \begin{tabular}[c]{@{}c@{}}2\\ (61)\end{tabular}   & \begin{tabular}[c]{@{}c@{}}2.2 mA\\ (75 $\upmu$A)\end{tabular}   & \begin{tabular}[c]{@{}c@{}}200 A/m\\ (6.9 A/m)\end{tabular}
\end{tabular}
\caption{ Performance comparison for the three integration methods, where $\delta B_{111}$, $\delta I$, and $\delta K$ are the magnetic field, current, and surface current density noise floors. Method 1 quotes the performance of Sample A and Methods 2 and 3 quote the performance for Sample B. The numbers in parentheses are projected for the more sensitive Sample A diamond if used for Methods 2 and  3. 
}
\label{schemesTable}
\end{table*}

\begin{figure*}[ht]
\begin{center}
\begin{overpic}[width=0.85\textwidth]{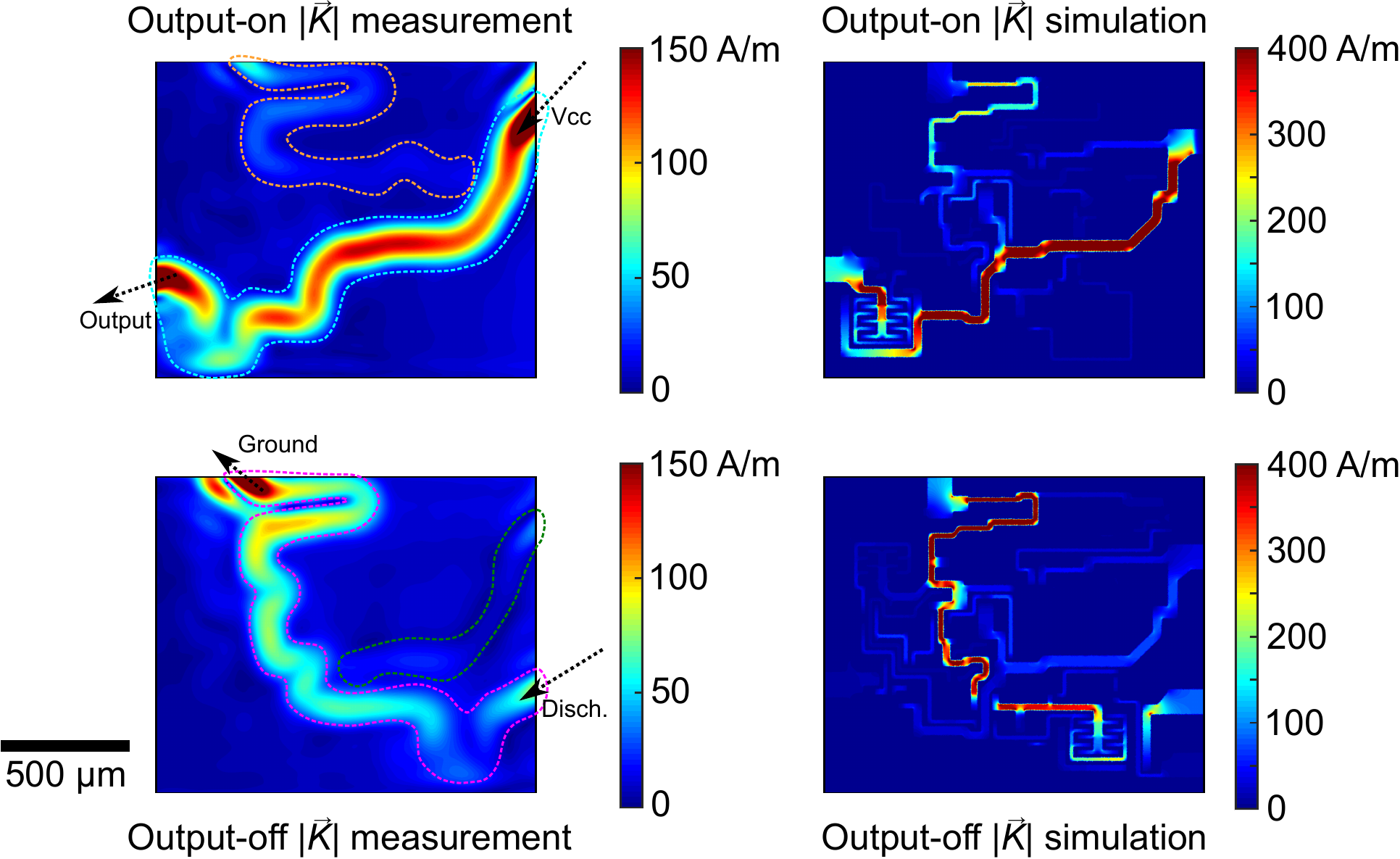}
\end{overpic}
\end{center}
\caption{\label{JxJyCOMSOL}
Output-on and output-off surface current density $|\vec{K}|$, from Method 1 measurements (Fig.~\ref{bigDiamondTop}) and from \texttt{COMSOL} simulation. The $|\vec{K}|$ maps calculated from the measurement have broadened features, since we suppress high spatial frequencies to suppress measurement noise. Here we show the simulated $|\vec{K}|$ for the metal layer only; the measurement can not tell the difference between current in the two conducting layers. The black arrows  show the primary current input/output points for the device, and the circled regions show current paths in Fig.~\ref{currentPaths} (with the same colors).
}
\end{figure*}

\section{Simulation and Measurement Analysis}

Figures \ref{bigDiamondTop}-\ref{smallDiamondTop} highlight magnetic features for the key current paths (Fig.~\ref{currentPaths}), which tell us about the 555 internal behavior in different states. For Method 1 (Fig.~\ref{bigDiamondTop}), the strongest magnetic field signatures come from current flowing from \texttt{Vcc} to \rload~(cyan) and as the capacitor discharges through the die to ground (magenta). In addition, we see weaker magnetic features as current flows through $\mathtt{R02}$ and $\mathtt{Q07}$ as part of a current mirror in the output-on state (orange) and as other components draw current seen in the output-off state (green). Although these features are coarse due to the standoff distance, these magnetic images can still be converted to surface current density maps (Fig.~\ref{JxJyCOMSOL}) that are consistent with those of Fig.~\ref{currentPaths}.  The Method 2 magnetic images (Fig.~\ref{backThinned}) also have magnetic features from the above current paths, though with the improved standoff distance, we also see hints (vertical stripes) of the 0.67 mA of current through  the three 5 k$\Omega$ resistors in series ($\mathtt{R07}$, $\mathtt{R08}$, and $\mathtt{R09}$, dark blue). For comparison, the Method 3 magnetic images (Fig.~\ref{smallDiamondTop}) also show current going to the 5 k$\Omega$ resistors, though the field from the resistors themselves is less prominent. Furthermore, this method has the most pronounced magnetic features for currents supplying additional components on the left side of the die (green), including for the output-on case. By analyzing these magnetic features, calculating the forward-model and inverse-model magnetic field and current density maps, and correlating these with the schematic and die layout, we confirm that the anticipated internal current paths are present for 555 states.

Using the $B_{111}$ measurements in Fig.~\ref{bigDiamondTop}, and approximating the 555 die currents as a 2D sheet, we calculated the surface current densities $\{K_x,K_y\}$. To do this, we used a Fourier analysis approach to compute the magnetic inverse problem needed to reconstruct the surface current densities from the measured $B_{111}$ maps \cite{wickswo89, degenCurrentImg, hollenbergCurrentImg, amirIEEE}. We used a  71 $\upmu$m standoff distance (known from the forward model), zero-padded the original magnetic map to help suppress edge artefacts, and applied a Hann filter in the frequency domain with $\lambda = 1.5 \times 71~\upmu$m cutoff wavelength to suppress measurement noise with high spatial frequencies.

Figure \ref{JxJyCOMSOL} shows the resulting $|\vec{K}|$ surface current amplitudes of this inverse-problem analysis. Comparing with the surface current densities calculated by the \texttt{COMSOL} simulation, we see good agreement, though the calculated $|\vec{K}|$ are broadened due to the standoff distance and the cutoff wavelength. The primary current paths (also drawn in Fig.~\ref{currentPaths}) give rise to the magnetic features highlighted in the measurements and simulations, and this inverse-problem analysis shows how we can correlate the magnetic image information with current paths in the 555 die.

\section{Conclusion and Outlook}
The QDM is a promising magnetic field diagnostic tool for integrated circuits that has micron-scale spatial resolution, millimeter-scale field of view area, and can operate at ambient conditions. In this paper we have appraised the performance of the QDM to measure fields from the die of a 555 timer using three measurement configurations and compared the measurement outputs to a finite element simulation. Comparing the experimental and computational outputs for the magnetic field shows good consistency, demonstrating that the QDM is capturing the expected magnetic field information available from the IC without being noticeably affected by artefacts or systematics \cite{tetienneArtefacts}. We also identified magnetic features and current paths in the measured and simulated results, confirming that we can glean accurate information about the internal current phenomena. Finally, since it has a few-micron feature size, up to 220 $\upmu$T magnetic field strength near the surface, and is feasilble (but also nontrivial) to simulate, the 555 is an ideal device with which to characterize and evaluate the QDM performance and techniques when used to sense electric currents in ICs.

These full-circuit simulation and measurement results establish a foundation from which to advance and optimize the state of the art of the QDM as a diagnostic tool for ICs. Continued work will explore how to classify measured magnetic maps using image analysis techniques (for example, using a structural similarity index measure or machine learning classification) to quickly identify working, faulty, and counterfeit ICs \cite{qdmFPGA}. Furthermore, since the QDM measures the magnetic fields from all pixels in parallel, this instrumentation can be extended to measuring IC dynamics as a magnetic movie, or can be modified to measure MHz- or GHz-frequency fields using NV AC magnetometry techniques \cite{linh_widefield, maletinskyMWimger}.

\section{Acknowledgements}
We thank Mark Gores (Integra Technologies) for help with die cross-section imaging and analysis, Dan Thompson for help with image preparation, George Burns for $^{15}$N ion implantation, and Shanalyn Kemme for help with optical microscope troubleshooting. Sandia National Laboratories is a multi-mission laboratory managed and operated by National Technology and Engineering Solutions of Sandia, LLC, a wholly owned subsidiary of Honeywell International, Inc., for the DOE's National Nuclear Security Administration under contract DE-NA0003525. This work was funded, in part, by the Laboratory Directed Research and Development Program and performed, in part, at the Center for Integrated Nanotechnologies, an Office of Science User Facility operated for the U.S.~Department of Energy (DOE) Office of Science.  This paper describes objective technical results and analysis. Any subjective views or opinions that might be expressed in the paper do not necessarily represent the views of the U.S. Department of Energy or the United States Government.  P.K.~is supported by the Sandia National Laboratories Truman Fellowship Program. E.V.L. is funded by the MITRE Corporation through the MITRE Innovation Program.

%\bibliography{NV555bibl}


\clearpage


\section*{Supplementary material}

# Supplemental material for "Measurement and Simulation of the Magnetic Fields from a 555 Timer Integrated Circuit using a Quantum Diamond Microscope and Finite Element Analysis"

P. Kehayias,[1, *] E. V. Levine,[2, 3, 4] L. Basso,[5] J. Henshaw,[5] M. Saleh Ziabari,[5, 6] M. Titze,[1] R. Haltli,[1] J. Okoro,[1] D. R. Tibbetts,[1] D. M. Udoni,[1] E. Bielejec,[1] M. P. Lilly,[5] T.-M. Lu,[1, 5] P. D. D. Schwindt,[1] and A. M. Mounce[5]

[1]*Sandia National Laboratories, Albuquerque, New Mexico 87185, USA*
[2]*The MITRE Corporation, Bedford, Massachusetts 01730, USA*
[3]*Quantum Technology Center, University of Maryland, College Park, MD 20742, USA*
[4]*Department of Physics, Harvard University, Cambridge, MA 02138, USA*
[5]*Center for Integrated Nanotechnologies, Sandia National Laboratories, Albuquerque, New Mexico 87123, USA*
[6]*University of New Mexico Department of Physics and Astronomy, Albuquerque, New Mexico 87131*

## 555 DIE CHARACTERIZATION

Figure S1 includes optical bright-field and dark-field microscope images of an example 555 die (RCA CA555CE [S1]). Each microscopy technique offers complementary information about the die layout and the doped-silicon layer, which we replicate in the COMSOL magnetic field simulation (Fig. 2b in the main text). Figure S2 labels the component locations shown in the PSPICE schematic (Fig. 3 in the main text).

To obtain the cross-section images of the 555 die (Fig. 2c-d in the main text), we extracted two dies from the packaging, then cross-sectioned them in planes along the vertical and horizontal directions. The dies were mounted onto pieces of raw silicon wafer and ground using an un-encapsulated cross-sectioning jig and standard metallographic processes, working down from 600-grit silicon carbide to a 0.04 µm silica slurry for polishing. The dies were then etched using a 10-10-1 nitric-acetic-hydrofluoric acid solution to delineate the interlayer dielectric layers, the metal layer, and doped-silicon layer diffusions. The samples were then inspected in an FE-SEM to document and measure the features of the device.

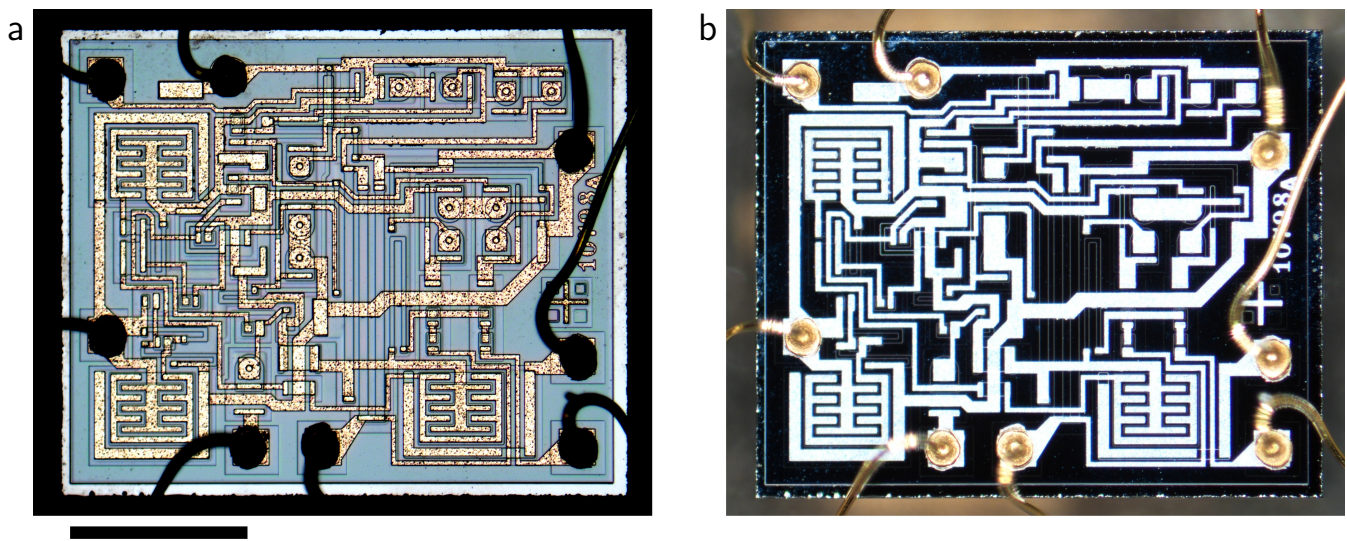

**Supplementary Figure** S1. (a) 555 die bright-field image. The metal-layer structures are yellow and grainy, the silicon-layer structures are slightly pink and outlined, and the bare silicon is gray. (b) 555 die dark-field image. The metal-layer structures are white, some silicon-layer structures are thinly outlined in white, and the bare silicon is black. Scalebar is 500 µm.

## PSPICE SIMULATION DETAILS

Figure 3 in the main text shows the schematic used to simulate the currents in the 555 device using the PSPICE circuit simulation software. The PSPICE output (voltages and currents for internal and external nodes and com-

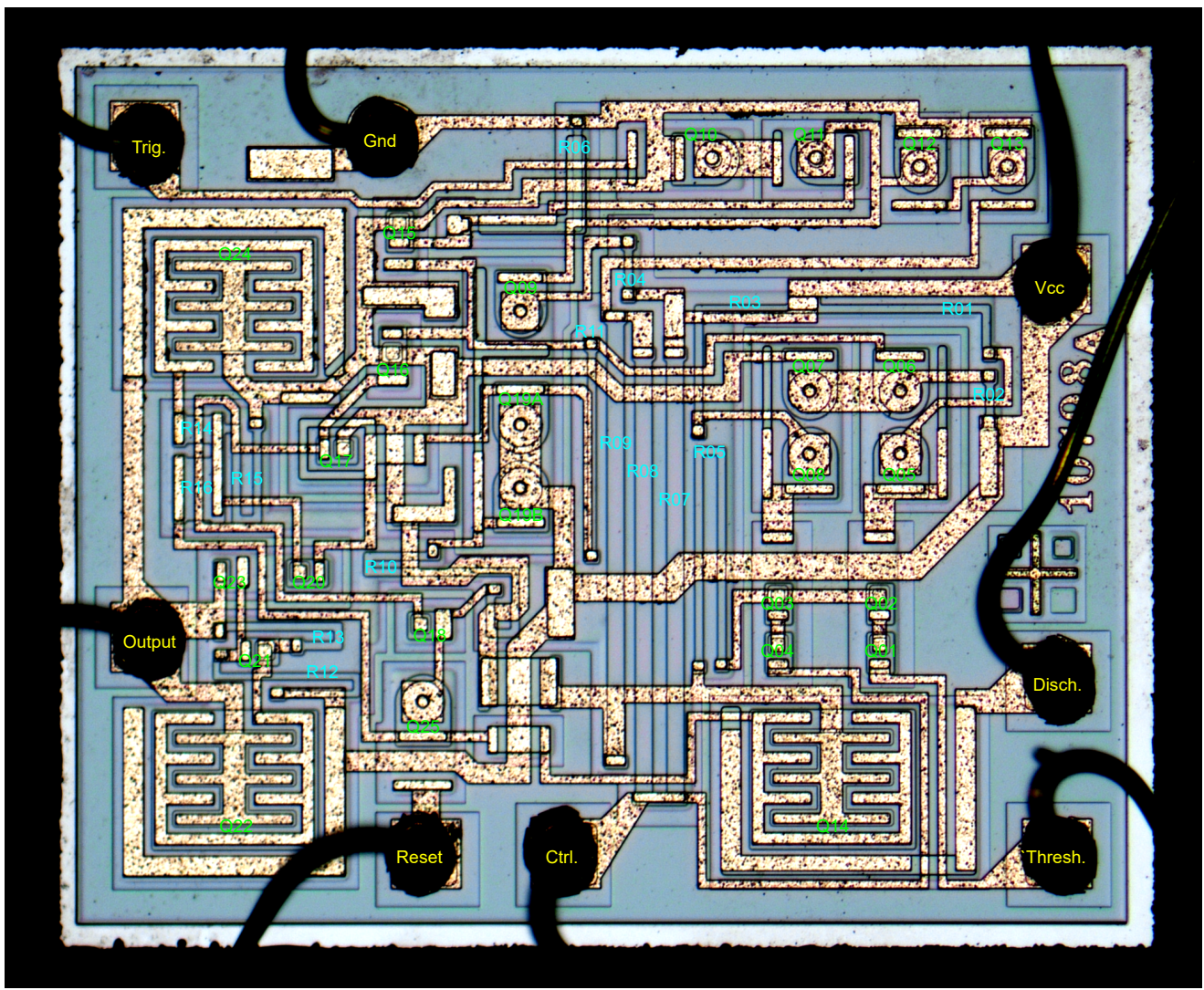

**Supplementary Figure** S2. Photo of the 555 die, with the components labeled as in the schematic (Fig. 3 in the main text).

ponents) were used in the `COMSOL` model to calculate the expected current distributions and magnetic field maps. This schematic combines the 555 internal schematic with the external components in the two-state oscillator demo circuit [S1–S3]. Note that we connected Pin 5 to ground through a 10 G$\Omega$ resistor in this schematic; we left this pin disconnected in the experiments. The measured external voltages and currents for the capacitor `C1` and the resistor `R`$_\text{load}$ are consistent with those predicted by `PSPICE`, as seen in Fig. S3.

## MAGNETIC IMAGING MEASUREMENT DETAILS

### Diamond sample properties

Table S1 includes further details for the diamond samples used in this work. Although Sample A has better properties for magnetic imaging with a low noise floor, it was unavailable for cutting and long-term use in this experiment, prompting us to make Sample B for some of the measurements.

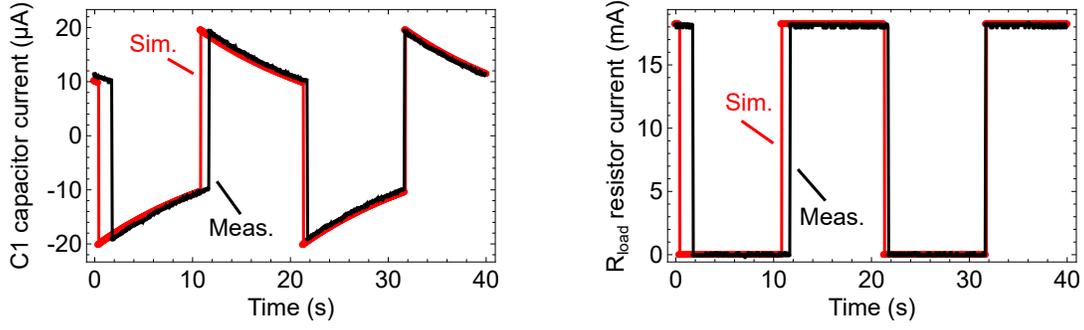

Supplementary Figure S3. Simulated currents through `C1` and `Rload`, compared to the currents measured with an oscilloscope.

| Sample | Dimensions | NV layer thickness | [N] in layer | [$^{13}$C] in layer | Lineshape FWHM | Fluor. contrast |
|---|---|---|---|---|---|---|
| A | 4×4×0.5 mm$^3$ | 4 μm | 20 ppm, $^{14}$N grown in | 0.001% | 1.1 MHz | 0.012 |
| B | Originally 4×4×0.5 mm$^3$, cut to 1.14×0.84×0.5 mm$^3$ | 1 μm | 50 ppm, $^{15}$N implant | 1.1% | 3.7 MHz | 0.0040 |

Supplementary Table S1. Relevant properties for the diamond samples used in this work.

### $\delta K$ and $\delta I$ noise floor conversions

To calculate the surface current density noise floor $\delta K$ from the magnetic noise floor $\delta B_{111}$ in Table 1 of the main text, we first assume that we are measuring the magnetic field from an infinite sheet of current in the $+x$ direction projected along the NV [111] crystallographic direction ($\{0, \sqrt{2/3}, \sqrt{1/3}\}$ in the lab frame). The NVs will detect a magnetic field of amplitude $B_{111} = \sqrt{2/3}\mu_0 K/2$, where $\mu_0 = 4\pi \times 10^{-7}$ m·T/A is the vacuum permeability and $K$ is the surface current density. This yields $\delta K = \frac{2}{\sqrt{2/3}\mu_0}\delta B_{111}$, which depends only on the magnetic noise floor and is independent of the standoff distance.

To calculate the current noise floor $\delta I$ from the magnetic noise floor $\delta B_{111}$, we first assume that we are measuring the magnetic field from a thin infinite wire in the $+x$ direction and crossing through $\{0, 0, -z_0\}$, where $z_0$ is the standoff distance. In the $x$-$y$ plane, the magnetic field is $\vec{B} = \{0, \frac{-\mu_0 I z_0}{2\pi(y^2+z_0^2)}, \frac{\mu_0 I y}{2\pi(y^2+z_0^2)}\}$, where $I$ is the current. The NVs will measure the field projection along the [111] direction, or $B_{111} = \frac{\mu_0 I(y-\sqrt{2}z_0)}{2\sqrt{3}\pi(y^2+z_0^2)}$. For a given $z_0$, the maximum field amplitude is $B_{111} = \frac{\mu_0 I}{(12-4\sqrt{6})\pi z_0}$. This yields $\delta I = \frac{(12-4\sqrt{6})\pi z_0}{\mu_0}\delta B_{111}$ as the current sensitivity noise floor with $z_0$ standoff distance.

### BACK-THINNED OUTPUT-ON MAGNETIC IMAGES WITH VARIABLE OUTPUT CURRENT

Figure S4 shows measurements of a back-thinned 555 die with different `Rload` values (470 Ω, 1.8 kΩ, and open-circuit) and output currents (18 mA, 4.8 mA, and 0 mA), connected through a rotary switch. This illustrates how the 555 internal behavior changes with the external components, and shows how these internal changes are detectable in the resulting magnetic images. Note that the 555 still draws current with 0 mA output current, which is visible as the current flows internally. As expected, the output-off magnetic image is unaffected when changing `Rload`.

### SIMULATED SURFACE CURRENT DENSITIES OF THE TWO CONDUCTING 555 DIE LAYERS

In Fig. S5, we compare the simulated surface current densities in the 555 metal layer and the doped-silicon layer. These show the current features highlighted in Fig. 3 of the main text, and we see that most of the surface current density is in the metal layer, though the doped-silicon layer also contributes. Note that the surface current densities

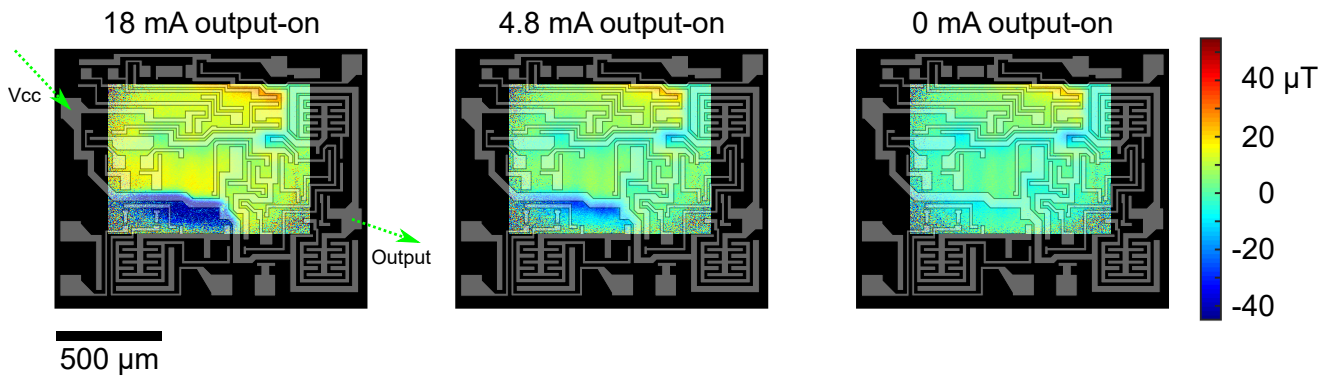

**Supplementary Figure** S4. Back-thinned output-on measurements for 18 mA (Fig. 6 in the main text), 4.8 mA, and 0 mA output currents. Reducing the output current mainly affects the trace in the lower-left corner, while the magnetic fields elsewhere are largely unaffected.

calculated from the NV magnetic microscopy measurements (Fig. 8 in the main text) do not differentiate between the two layers, returning the sum of the two layers.

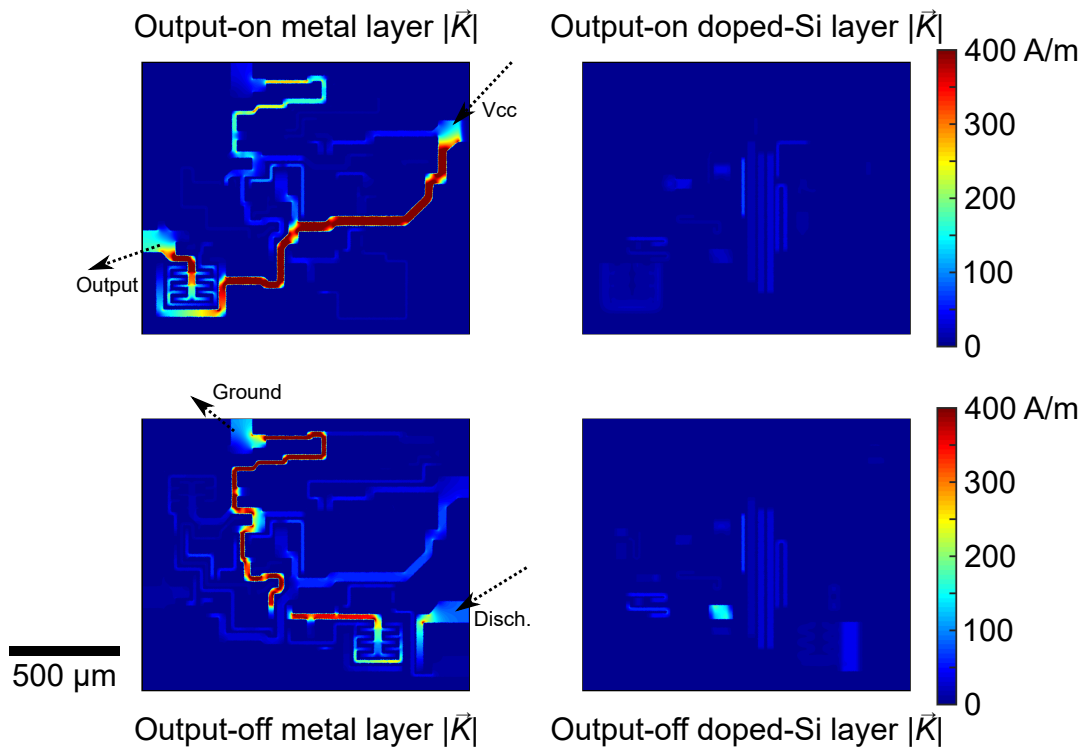

**Supplementary Figure** S5. Comparison between simulated surface current densities in the 555 metal layer and in the doped-silicon layer.



\end{document}